\documentclass[twocolumn]{aastex631}

\hypersetup{linkcolor=magenta,citecolor=blue,filecolor=cyan,urlcolor=purple}

\usepackage{graphicx}
\usepackage[caption=false]{subfig}
\usepackage{booktabs}
\usepackage{amsmath}
\usepackage[thinc]{esdiff}

\usepackage{rotating}

\received{July 28, 2023}
\revised{September 18, 2023}
\accepted{October 9, 2023}
\submitjournal{ApJ}

\shorttitle{Interplanetary Propagation of a Large Filament Eruption}
\shortauthors{Palmerio et al.}

\graphicspath{{./}{figures/}}

\turnoffeditone
\turnoffedittwo
\turnoffeditthree

\begin{document}

\title{Modeling a Coronal Mass Ejection from an Extended Filament Channel. \\ II. Interplanetary Propagation to 1 au}

\correspondingauthor{Erika Palmerio}
\email{epalmerio@predsci.com}

\author[0000-0001-6590-3479]{Erika Palmerio}
\affiliation{Predictive Science Inc., San Diego, CA 92121, USA}

\author[0000-0002-4269-056X]{Anwesha Maharana}
\affiliation{Centre for mathematical Plasma Astrophysics (CmPA), KU Leuven, 3001 Leuven, Belgium}
\affiliation{Solar--Terrestrial Centre of Excellence---SIDC, Royal Observatory of Belgium, 1180 Brussels, Belgium}

\author[0000-0001-6886-855X]{Benjamin J. Lynch}
\affiliation{Space Sciences Laboratory, University of California--Berkeley, Berkeley, CA 94720, USA}

\author[0000-0002-5681-0526]{Camilla Scolini}
\affiliation{Space Science Center, University of New Hampshire, Durham, NH 03824, USA}

\author[0000-0002-4921-4208]{Simon W. Good}
\affiliation{Department of Physics, University of Helsinki, FI-00014 Helsinki, Finland}

\author[0000-0003-1175-7124]{Jens Pomoell}
\affiliation{Department of Physics, University of Helsinki, FI-00014 Helsinki, Finland}

\author[0000-0002-7178-627X]{Alexey Isavnin}
\affiliation{Rays of Space Oy, FI-00750 Helsinki, Finland}

\author[0000-0002-4489-8073]{Emilia K. J. Kilpua}
\affiliation{Department of Physics, University of Helsinki, FI-00014 Helsinki, Finland}

% ===== ABSTRACT =====

\begin{abstract}

We present observations and modeling results of the propagation and impact at Earth of a high-latitude, extended filament channel eruption that commenced on 2015 July 9. The coronal mass ejection (CME) that resulted from the filament eruption was associated with a moderate disturbance at Earth. This event could be classified as a so-called ``problem storm'' because it lacked the usual solar signatures that are characteristic of large, energetic, Earth-directed CMEs that often result in significant geoeffective impacts. We use solar observations to constrain the initial parameters and therefore to model the propagation of the 2015 July 9 eruption from the solar corona up to Earth using 3D magnetohydrodynamic heliospheric simulations with three different configurations of the modeled CME. We find the best match between observed and modeled arrival at Earth for the simulation run that features a toroidal flux rope structure of the CME ejecta, but caution that different approaches may be more or less useful depending on the CME--observer geometry when evaluating the space weather impact of eruptions that are extreme in terms of their large size and high degree of asymmetry. \edit1{We discuss our results in the context of both advancing our understanding of the physics of CME evolution and future improvements to space weather forecasting.}

\end{abstract}


% ===== INTRODUCTION =====

\section{Introduction} \label{sec:intro}

Geoeffective interplanetary coronal mass ejections (ICMEs) that have no clear solar counterpart or that are associated with ambiguous or negligible solar activity are usually referred to as ``problem storms.'' The concept was first introduced by \citet{dodson1964}, who surveyed a large number of geomagnetic storms and could not always find a preceding solar event to account for the disturbances at Earth. It should be noted, however, that in the 1960s the sources of the largest geomagnetic disturbances used to be searched uniquely in solar flares \citep{gosling1993}. With the first detections of coronal mass ejections (CMEs) in the 1970s using coronagraph observations, the role of solar eruptions in causing geomagnetic activity at Earth became increasingly evident. Alongside flares, it was noted that possible low-coronal signatures that indicate the occurrence of a CME eruption include post-eruption arcades, coronal dimmings, and disappearing filaments \citep[e.g.,][]{hudson2001}. Nevertheless, the problem storm issue still survives to this day \citep[e.g.,][]{nitta2021}.

What makes a CME ``problematic''? In principle, any geomagnetic storm that is (fully or mostly) unexpected can be defined as a problem storm. Besides the well-known, current issues in space weather forecasting, related e.g.\ to CME evolution and interactions in interplanetary space \citep[e.g.,][]{kilpua2019a,vourlidas2019}, additional complexity may arise from CMEs that lack low-coronal signatures \citep[so-called ``stealth CMEs;'' e.g.,][]{robbrecht2009b,nitta2017}, that are not visible in coronagraph imagery \citep[e.g.,][]{howard2008b,palmerio2019}, and/or that do not present obvious Earth-directed components \citep[e.g.,][]{schwenn2005,kilpua2014}. Such CMEs may feature diverse characteristics, but what makes them problematic is that their potential geoeffectiveness usually goes unnoticed when formulating space weather forecasts.

One interesting class of CMEs in this regard is that of streamer-blowout CMEs \citep[e.g.,][]{sheeley1982,vourlidas2018}, which are large-scale, gradual events that originate from the quiet Sun and that are followed by wide and slow CMEs. Streamer blowouts are not accompanied by flares because of their gradual and slow nature, as well as their usual initiation sites higher up in the corona, thus they are generally not considered as potentially geoeffective as the more impulsive active-region CMEs. An excellent example of a geoeffective streamer-blowout CME, however, was presented by \citet{mcallister1996}, who reported a CME with a large source region (spanning ${\sim}150^{\circ}$ in longitude and ${\sim}35^{\circ}$ in latitude) that could be identified only because of the appearance of an extended post-eruption arcade in soft X-ray data. The lack of an associated flare or filament disappearance made the following severe geomagnetic storm largely unexpected.

In this paper, we build upon the work of \citet{lynch2021}, hereafter Paper~I, to analyze the propagation of an extended, high-latitude filament eruption up to Earth. The event that we focus on commenced on 2015 July 9 and is in several aspects analogous to the CME reported by \citet{mcallister1996} in the sense that, despite affecting a significant portion of the solar southern hemisphere upon eruption, both events were characterized by faint, non-impulsive on-disk signatures, characteristic of a streamer-blowout CME. Furthermore, both CMEs caused problem storms at Earth, although with different magnitudes \citep[Dst$_\text{min}$ = $-203$~nT for the][event and Dst$_\text{min}$ = $-68$~nT for the event described in this paper]{mcallister1996}.

While the effort of Paper~I was centered on modeling the eruption and initial evolution (up to 30\,$R_{\odot}$) of the CME, in this work we focus on modeling the propagation of the CME from the solar corona (21.5\,$R_{\odot}$ or 0.1~au) up to Earth's orbit (1~au) using magnetohydrodynamic (MHD) simulations. Due to the intrinsic complexity of simulating such a large-scale structure, we initialize our CME inputs using three different configurations of the modeled ejecta: a hydrodynamic pulse, a spheromak, and a toroidal flux rope. Our aim is to investigate the impact that the CME's internal magnetic structure and configuration has on the ejecta's heliospheric evolution and synthetic in-situ profiles, in order to \edit1{improve our understanding of complex ``global'' eruptions that, despite their faint and slow nature, have the potential to drive geomagnetic effects and, as a consequence, to} best inform future space weather forecasting and modeling.

This paper is organized as follows. Section~\ref{sec:observations} provides an observational overview of the 2015 July~9 CME from the Sun to Earth, starting from the eruption itself analyzed in detail in Paper~I ($\S$\ref{subsec:eruption}), following the CME propagation through the solar corona ($\S$\ref{subsec:corona}), and finally observing the corresponding ICME at Earth's Lagrange L1 point ($\S$\ref{subsec:earth}). Section~\ref{sec:setup} presents the modeling setup for the three simulations, first in terms of the solar wind background ($\S$\ref{subsec:background}) and then in terms of the CME input parameters for the three different ejecta configurations ($\S$\ref{subsec:modelcme}). Section~\ref{sec:results} shows our modeling results aimed at connecting the CME structure in the corona to its associated ICME near Earth, including a comparison of the resulting CMEs in the heliosphere with the simulated CME in the corona from Paper~I ($\S$\ref{subsec:simresults}) and the corresponding time series at Earth ($\S$\ref{subsec:simearth}). In Section~\ref{sec:discussion} we discuss our results in the context of predicting the space weather effects of large and slow CMEs. Finally, in Section~\ref{sec:conclusions} we summarize the results and present our conclusions.

% ===== OBSERVATIONS =====

\section{Overview of the Observations} \label{sec:observations}

The onset and eruption of the 2015~July~9 CME was analyzed and modeled in detail in Paper~I, hence we provide only a brief review in Section~\ref{subsec:eruption}. We discuss the evolution of the CME through the solar corona based on corongraph observations in Section~\ref{subsec:corona}. Finally, we present the measurements and analysis of the corresponding ICME near Earth in Section~\ref{subsec:earth}.

% ===== RECAP OF PAPER I =====

\subsection{CME Eruption at the Sun} \label{subsec:eruption}

%%%%% FIGURE 1 %%%%%
\begin{figure*}[!t]
\center
	\includegraphics[width=0.99\linewidth]{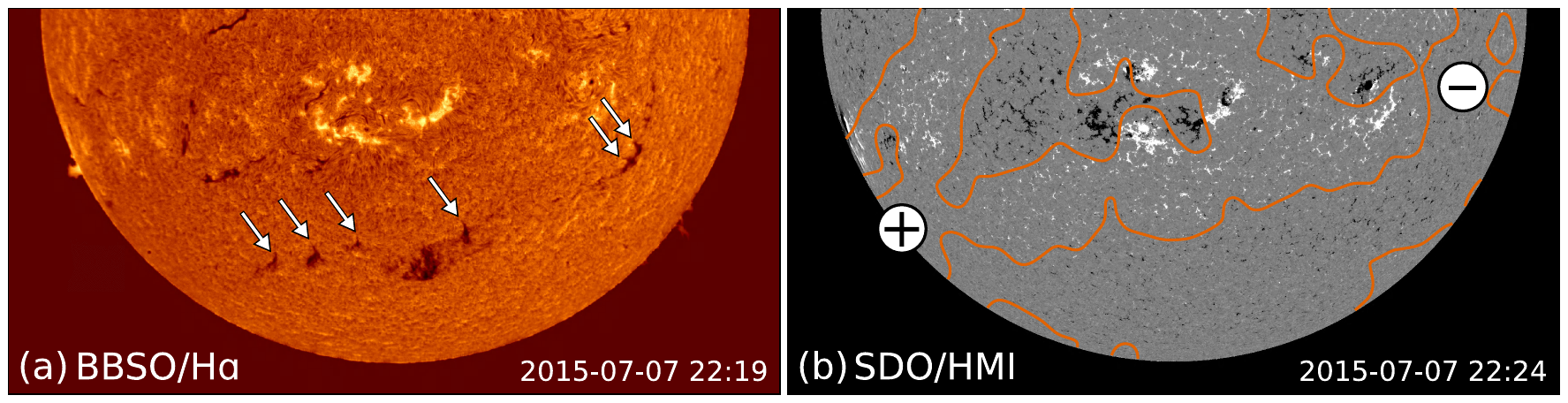}
	\caption{Pre-eruptive configuration of the filament. (a) H$\alpha$ image from BBSO, with left-bearing filament barbs (signatures of a right-handed filament) indicated by arrows. (b) SDO/HMI magnetogram with the global PILs marked in orange. The approximate locations of the positive and negative filament footpoints with respect to the PIL involved in the eruption are marked with $\oplus$ and $\ominus$ symbols, respectively.}
	\label{fig:disc}
\end{figure*}
%%%%%%%%%%%%%%%%%%%

The eruption of the CME analyzed in this work initiated on 2015 July 9 around 19:00~UT. Solar extreme ultra-violet (EUV) observations from the Atmospheric Imaging Assembly \citep[AIA;][]{lemen2012} instrument on board the Solar Dynamics Observatory \citep[SDO;][]{pesnell2012} reveal a large southern-hemisphere filament spanning the whole Earth-facing disk in longitude and slowly erupting asymmetrically from its eastern leg toward its western one. The whole eruption occurred over large time scales, taking approximately 12~hr for the filament to completely disappear from the AIA field of view (see Figure~1 and the accompanying animation in Paper~I).

To estimate the magnetic properties of the CME at the time of the eruption, or intrinsic flux rope type \citep[see][and references therein]{palmerio2017}, we combine chromospheric observations of the pre-eruptive filament with photospheric measurements of the solar magnetic field, both shown in Figure~\ref{fig:disc}. Respectively, we use data from the Big Bear Solar Observatory (BBSO) full-disk H$\alpha$ telescope \citep[][]{denker1999} and the Helioseismic and Magnetic Imager \citep[HMI;][]{scherrer2012} on board SDO. In the case of large quiet-Sun filaments, such as the one involved in this eruption, the study of filament barbs, fine structures seen along the sides of filament spines, is an excellent method for determining the corresponding magnetic helicity sign or chirality \citep[e.g.,][]{martin2003}. \edit1{Barbs are interpreted as appendages departing from the main filament body and extending down to the chromosphere, and their orientation with respect to the spine provides information as to the sense of twist of the embedded magnetic field \citep[e.g.,][]{pevtsov2003b}.} The filament shown in Figure~\ref{fig:disc}(a) is characterized by left-bearing barbs, a signature of sinistral filaments and thus indicating a right-handed flux rope. The magnetogram data shown in Figure~\ref{fig:disc}(b) confirm that the source of the eruption corresponds to an extended quiet-Sun filament channel, and the assumption of right-handed chirality implies that the filament's eastern (western) footpoint is rooted in the positive (negative) polarity on either side of the underlying polarity inversion line (PIL), \edit1{realizing an overall forward-S shape \citep[e.g.,][]{rust2003}}. Then, the flux rope axial field is simply expected to run from the positive polarity to the negative one, i.e.\ toward the west in the case of this event.

From the analysis of the pre-eruptive structure based on the data shown in Figure~\ref{fig:disc}, we can conclude that the 2015 July 9 CME flux rope was characterized by a right-handed chirality (or positive helicity), as well as that its axis had a low inclination to the solar equator (note how both the filament and the underlying PIL span the solar disk in the east--west direction, with an inclination of ${\sim}20^{\circ}$ to the equatorial plane) and was directed toward the west. This suggests a south--west--north (SWN) flux rope type at the time of the eruption, following the nomenclature of \citet{bothmer1998} and \citet{mulligan1998}.

% ===== CORONAL EVOLUTION =====

\subsection{CME Evolution in the Corona} \label{subsec:corona}

After its eruption from the Sun, we follow the initial propagation of the 2015~July~9 CME in the corona using the C2 and C3 telescopes, which are part of the Large Angle and Spectrometric Coronagraph \citep[LASCO;][]{brueckner1995} on board the Solar and Heliospheric Observatory \citep[SOHO;][]{domingo1995}. The east-to-west asymmetry that can be seen in EUV observations (see Section~\ref{subsec:eruption}) is well reflected in white-light ones. The CME is first observed in the LASCO/C2 field of view around 20:00~UT on July~9 as a relatively narrow streamer blowout to the southeast of the solar disk. A few hours later, around 03:00~UT on July~10, this structure is followed by an extended loop-like feature that sweeps the whole southern hemisphere in an asymmetric way, from east to west. Finally, around 14:30~UT on July~10, the loop is followed by a structure to the southwest that is reminiscent of a three-part CME cavity. The same features (streamer blowout to the southeast, large loop structure to the south, three-part cavity to the southwest) can also be discerned in LASCO/C3 imagery, starting around 02:00~UT on July~10. Overall, the passage of the whole 2015 July 9 CME through the two coronagraphs' fields of view takes place over $\sim$1.5~days, largely due to the multi-part structure of the eruption and to the ejecta's slow propagation speed (see Figure~2 and the accompanying animation in Paper~I).

We note that in the Computer Aided CME Tracking \citep[CACTus;][]{robbrecht2004, robbrecht2009a} catalog, the eruption under study is reported as two separate CMEs, the first appearing on 2015 July 9 at 20:00~UT (corresponding to the initial streamer blowout) and the second appearing on 2015 July 10 at 02:48~UT (corresponding to the loop-like ejected material). In the SOHO/LASCO CME catalog \citep{yashiro2004, gopalswamy2009b}, the CME is also reported as two separate partial halos, one starting on 2015 July 9 at 19:00~UT and the second starting on 2015 July 10 at 02:24~UT. We remark that the CACTus catalog is algorithm-generated (i.e., without human supervision) while the SOHO/LASCO one is manually generated, hence these results highlight the complexity of the event under study.

%%%%% FIGURE 2 %%%%%
\begin{figure}[!t]
\center
	\includegraphics[width=0.99\linewidth]{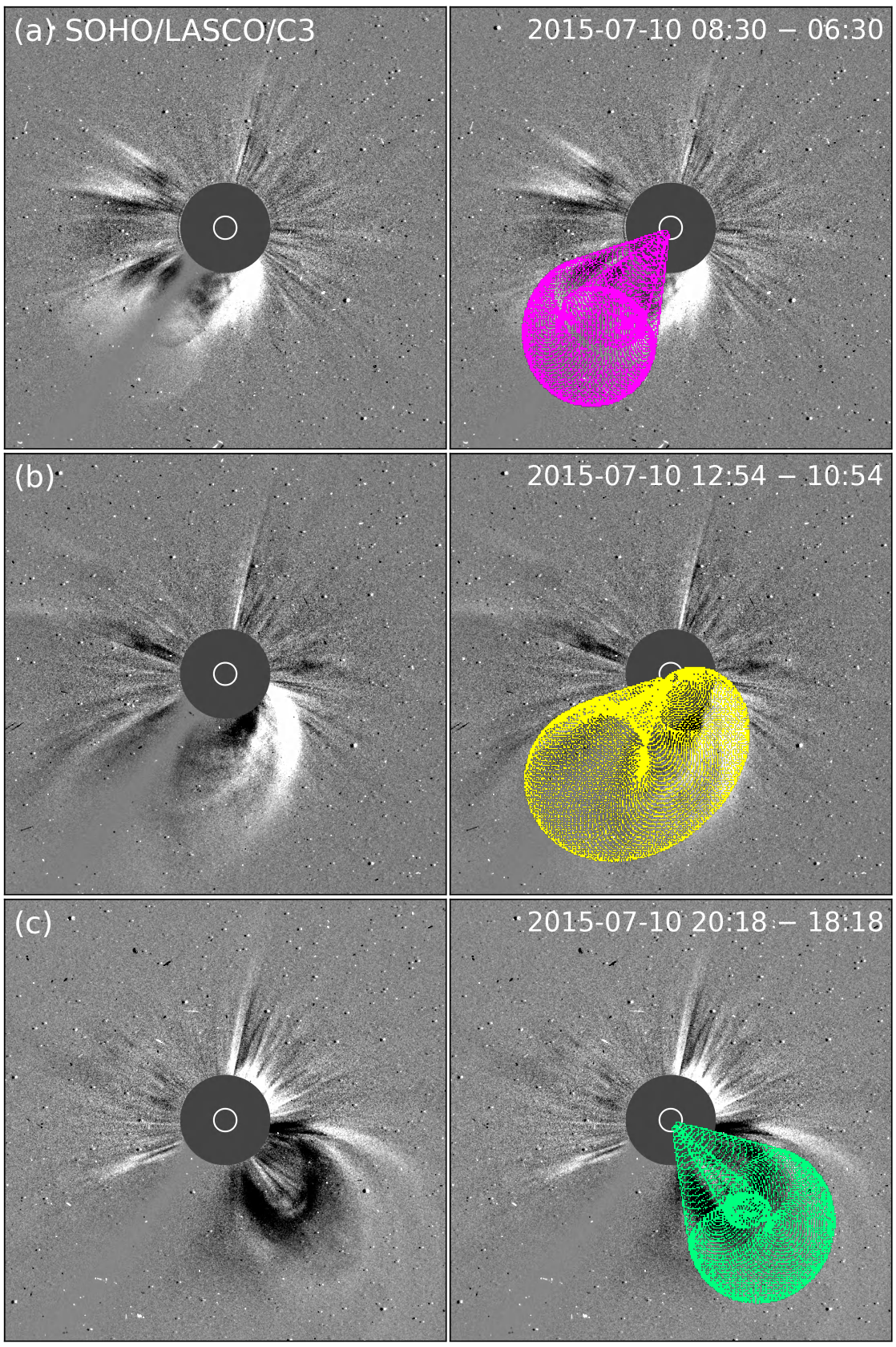}
	\caption{GCS reconstructions of the three different sub-parts---(a) streamer blowout, (b) large loop, and (c) three-part cavity---of the complex 2015 July 9 CME using the single LASCO/C3 viewpoint. The panels to the left show 2-hour difference images and the panels to the right display the corresponding GCS wireframe projection overlaid.}
	\label{fig:gcs_fits}
\end{figure}
%%%%%%%%%%%%%%%%%%%

To obtain an estimate of the CME's geometric and kinematic parameters, we perform reconstructions of its structure based on coronagraph images using the Graduated Cylindrical Shell \citep[GCS;][]{thernisien2011} model, which consists of a parameterized shell (described by six parameters) that is meant to reproduce the morphology of CMEs in the corona. By projecting the GCS wireframe onto the plane of sky of coronagraph imagery, a user can adjust the six free parameters until the obtained shell visually matches the CME's appearance. Forward modeling with the GCS technique is often performed to derive CME input parameters for \edit1{analytical \citep[e.g.,][]{kay2017c, calogovic2021} and MHD \citep[e.g.,][]{scolini2020, palmerio2022a} simulations}, but in the case of the 2015 July 9 event there are two factors that make the analysis more complex than usual. First, the Solar Terrestrial Relations Observatory Ahead (STEREO-A) spacecraft was in superior conjunction with Earth, meaning that most of its instruments were not operational and only the SOHO (Earth) vantage point is available. Although forward modeling with the GCS is usually performed with simultaneous coronagraph images from two or three viewpoints \citep{thernisien2009}, the technique was initially developed and tested using uniquely the SOHO/LASCO perspective \citep{thernisien2006}. One caveat to keep in mind, however, is that the uncertainties stemming from a single viewpoint will be inevitably larger \citep{verbeke2022}. Second, the strongly-asymmetric and slow nature of the 2015 July 9 event results in the CME never appearing in its entirety in a single coronagraph image. Thus, to take all the different sub-parts of the erupting CME into account, we perform three separate GCS fits for the streamer blowout, the large loop structure, and the three-part cavity, respectively.

%%%%% TABLE 1 %%%%%
\begin{table}[!t]
\centering
\caption{GCS parameters for each of the three sub-parts of the 2015 July 9 CME shown in Figure~\ref{fig:gcs_fits}. Latitude and longitude are reported in Stonyhurst coordinates. The tilt is measured from the solar west direction and is defined positive for counterclockwise rotations. The half-width is the half-angular distance between the axes of the CME legs. The aspect ratio is the ratio of the CME size at two orthogonal directions.
\label{tab:gcs}}
\hspace*{-0.18\columnwidth}
\begin{tabular}{lccc}
\toprule
GCS parameter & Part 1 & Part 2 & Part 3 \\[0.02in]
\midrule
Latitude ($\theta$)  & $-33^{\circ}$ &  $-40^{\circ}$  & $-35^{\circ}$ \\
Longitude ($\phi$) & $-32^{\circ}$ &  $-22^{\circ}$ & $40^{\circ}$ \\
Axial tilt ($\gamma$) & $90^{\circ}$ &  $23^{\circ}$ & $70^{\circ}$ \\
Nose height ($H$) & $21.5\,R_{\odot}$ & $21.5\,R_{\odot}$ & $21.5\,R_{\odot}$\\
Half-width ($\alpha$) & $6^{\circ}$ &  $43^{\circ}$ & $10^{\circ}$ \\
Apect ratio ($\kappa$) & 0.33 & 0.43 & 0.30 \\[0.02in]
\bottomrule
\end{tabular}
\end{table}
%%%%%%%%%%%%%%%%%%%

Snapshots of the different CME features through the SOHO/LASCO/C3 field of view together with the corresponding GCS fits are shown in Figure~\ref{fig:gcs_fits}, and the GCS parameters obtained for each reconstruction are presented in Table~\ref{tab:gcs}. It is clear from these results that all sub-parts of the 2015 July 9 CME propagate strongly toward the south, with latitudinal directions of the apexes lying around $-35^{\circ}$. It is also evident that so-called `Part~1' and `Part~3' of the event are directed away from the Sun--Earth line, while from `Part~2' it is possible to expect a grazing encounter at Earth with the CME's northern flank. 

% ===== IN-SITU OBSERVATIONS =====

\subsection{CME Measurements at Earth} \label{subsec:earth}

%%%%% FIGURE 3 %%%%%
\begin{figure}[!t]
\center
	\includegraphics[width=0.99\linewidth]{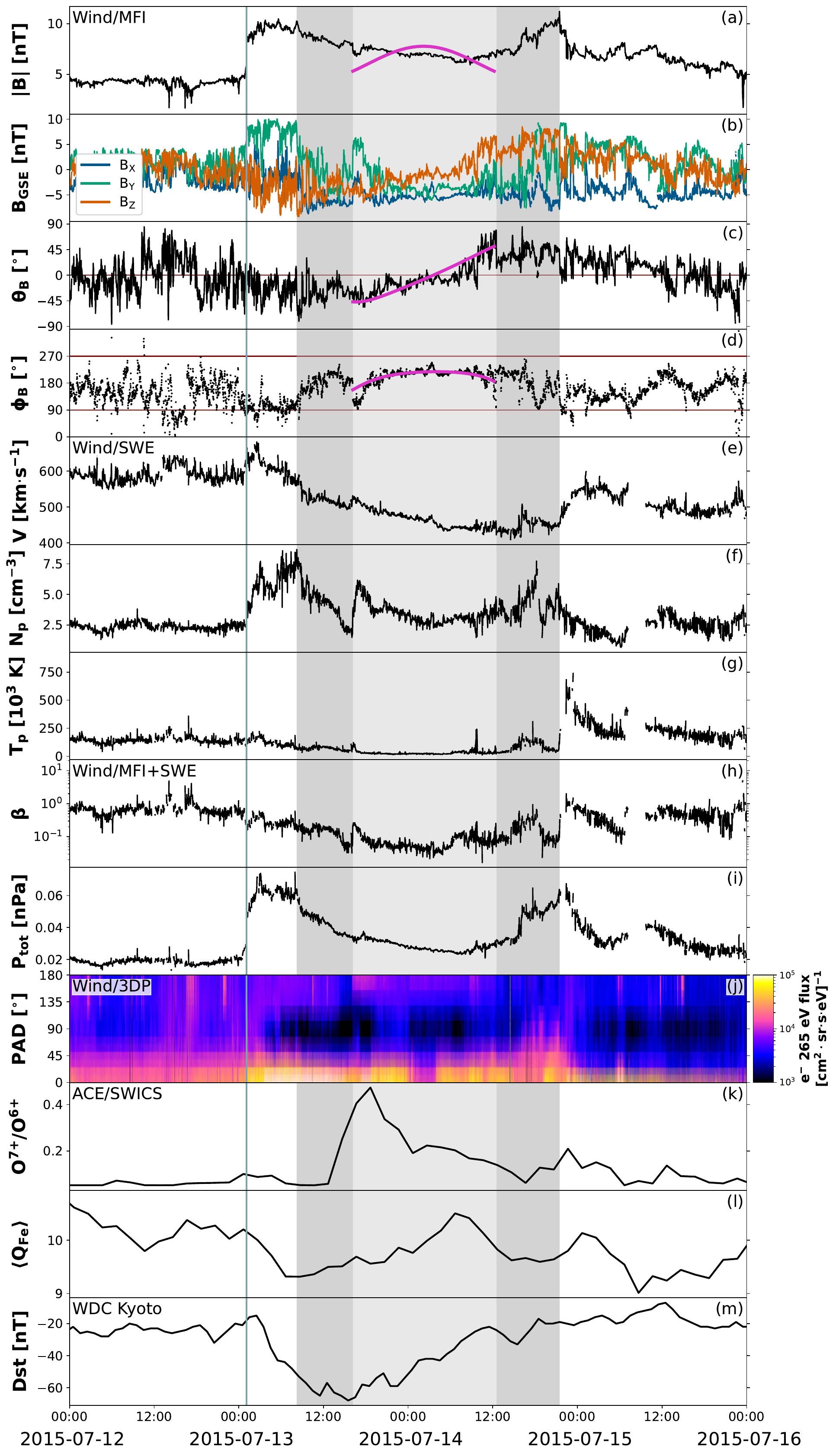}
	\caption{The 2015~July~9 CME observed in situ by the Wind and ACE spacecraft, together with its geomagnetic response. The panels show: (a) magnetic field magnitude, (b) Cartesian components of the magnetic field, (c) latitudinal and (d) longitudinal angles of the magnetic field, (e) solar wind speed, (f) proton density, (g) proton temperature, (h) plasma beta, (i) total pressure (proton and electron gas pressure plus magnetic pressure), (j) pitch-angle spectrogram of suprathermal 265~eV electrons, (k) oxygen charge state ratio, (l) average iron charge state, and (m) Dst index. The gray vertical line indicates the arrival of the shock-like disturbance, while the gray-shaded areas highlight the ICME ejecta (dark gray) and the magnetic cloud interval (light gray) within it. A Lundquist flux rope fitting has been plotted in pink over the magnetic cloud interval.}
	\label{fig:insitu}
\end{figure}
%%%%%%%%%%%%%%%%%%%

Despite propagating considerably toward the south (as shown from the GCS estimates presented in Section~\ref{subsec:corona}), the 2015 July 9 CME impacted Earth on July 13. Figure~\ref{fig:insitu} shows measurements taken from Earth's Lagrange L1 point by the Wind \citep{ogilvie1997} and Advanced Composition Explorer \citep[ACE;][]{stone1998} spacecraft, together with Dst index values, during the days following the July 9 eruption. The spacecraft data are provided by the Magnetic Field Investigation \citep[MFI;][]{lepping1995}, Solar Wind Experiment \citep[SWE;][]{ogilvie1995}, and Three-Dimensional Plasma and Energetic Particle Investigation \citep[3DP;][]{lin1995} instruments on board Wind as well as the Solar Wind Ion Composition Spectrometer \citep[SWICS;][]{gloeckler1998} on board ACE. The final Dst index is supplied by the World Data Center (WDC) for Geomagnetism, Kyoto.

The sequence of disturbances commences with a shock-like feature around 01:00~UT on 2015 July 13. This structure cannot be defined as a fully developed shock, since a clear jump is observed only in the magnetic field magnitude, plasma density, and total pressure, although the solar wind bulk speed experiences a small enhancement. Nevertheless, this feature marks the arrival of the CME-driven sheath region, and a full shock is likely not formed because the CME is traveling in a stream of fast solar wind---the plasma speed lies around 600~km$\cdot$s$^{-1}$ before the arrival of the disturbance. Starting around 08:00~UT on July 13, the ICME ejecta is seen to pass by the observing spacecraft. The ejecta shows clear ICME signatures \citep[e.g.,][]{zurbuchen2006}, including a declining speed profile (indicating expansion) and bidirectional electrons (indicating a structure that is still connected to the Sun from both legs). Additionally, the ejecta features magnetic cloud signatures \citep[e.g.,][]{burlaga1981}, including an elevated magnetic field magnitude, smoothly rotating magnetic field vectors, as well as a low temperature and plasma beta. The ICME ejecta boundaries, however, seem to extend beyond the magnetic cloud boundaries, leading to a five-part ICME \citep{kilpua2013} that includes a shock, a sheath region, an ICME front region, a magnetic cloud, and an ICME rear region\edit1{---each marked with different lines or shaded areas in Figure~\ref{fig:insitu}}.

This aspect is also evident from charge state data, which display more elevated values inside the magnetic cloud region than within the extended ICME ejecta region. \edit1{The O$^{7+}$/O$^{6+}$ charge state profile shows a distinct plasma population during the magnetic ejecta interval, whereas the $\langle Q_{\rm Fe}\rangle$ profile is more ambiguous. However, we note that the values are less enhanced than the thresholds typically used to identify ICME material: O$^{7+}$/O$^{6+} \gtrsim 1.0$ \citep{reinard2001} and $\langle Q_\mathrm{Fe}\rangle \gtrsim 11$ \citep{lepri2004}. In fact, both composition quantities are comfortably within ranges typically associated with the slow solar wind from helmet streamer and pseudostreamer source regions \citep[see, e.g.,][and references therein]{zhao2017, lynch2023}. These composition signatures are entirely consistent with stealth and slow streamer-blowout CMEs that tend to originate from larger coronal heights and are significantly less energetic than fast CMEs from active regions. Additional energization (heating) of the CME plasma due to interaction with the ambient solar wind in the heliosphere is unlikely to have any significant impact on these composition quantities given that the ``freeze-in'' distances for the relevant oxygen and iron charge states experiencing CME-like evolution are on the order of ${\lesssim}4\,R_\odot$ and ${\lesssim}10\,R_\odot$, respectively \citep{rivera2019}.} Finally, as mentioned in the Introduction, this event reached Dst$_\text{min}$ = $-68$~nT, corresponding to a moderate geomagnetic storm, possibly induced by the turbulent sheath region and the initial portion of ejecta material, since the index peaked approximately at the \edit1{front boundary} of the magnetic cloud interval.

Another interesting feature of this in-situ event is the magnetic field magnitude (and total pressure) profile, which peaks at the very front and very rear of the whole ICME structure. This trend appears remarkably similar to a series of CMEs observed in situ by Ulysses at ${\sim}3.5$~au and ${\sim}60^{\circ}$ latitude and described by \citet{gosling1994}. The ICMEs detected by Ulysses were bounded by a forward--reverse shock pair, possibly driven by CME overexpansion in interplanetary space rather than by an intrinsic high radial speed. Interestingly, one of these Ulysses ICMEs was the interplanetary counterpart of the large, high-latitude filament eruption reported by \citet{mcallister1996}. In the case of the event investigated here, we note that the discontinuity at the end of the structure cannot be defined as a reverse shock, since the temperature profile features an increase rather than a decrease. It is possible that the structure encountered in 2015 July was not characterized by a forward--reverse shock pair because it was detected at 1~au rather than at ${\sim}3.5$~au and along the ecliptic plane (yielding a flank encounter) rather than at high latitudes (resulting in a more central impact). Alternatively, the compression observed toward the rear of the ICME ejecta may be uniquely due to the faster solar wind following it, suggesting that the CME was embedded in a high-speed stream from either side---a rather unusual occurrence that has been, nevertheless, reported in the literature \citep[e.g.,][]{heinemann2019}. SDO/AIA observations in the 211~{\AA} channel (not shown), in fact, reveal the presence of a single coronal hole that could have been responsible for the fast solar wind measured at Earth (located along the solar equator and with a longitude of ${\sim}35^{\circ}$ west on 2015 July 9).

To evaluate the flux rope properties of the 2015 July 9 ICME, we perform a Lundquist force-free flux rope fitting \citep{lundquist1950,burlaga1988,lepping1990}, using the procedure described in \citet{good2019}, over the interval that we identified as a magnetic cloud (i.e., the light-gray shaded area in Figure~\ref{fig:insitu}). The method yields a right-handed flux rope with normalized impact parameter $p=0.64$ and axis orientation ($\vartheta_{0}$, $\varphi_{0}$) = ($-12^{\circ}$, $269^{\circ}$). Additionally, a Lundquist fitting performed over the whole interval that we identified as ICME ejecta (i.e., including the dark-gray shaded areas in Figure~\ref{fig:insitu}) reports a right-handed flux rope as well, with $p=0.70$ and ($\vartheta_{0}$, $\varphi_{0}$) = ($-11^{\circ}$, $264^{\circ}$). Thus, different fitting boundaries consistently yield a right-handed, low-inclination flux rope with a westward axis that is encountered fairly far from its center. This is consistent with the hypothesis of the 2015 July 9 CME propagating mainly toward the south and encountering Earth only through its northern edge.

% ===== MODELLING SETUP =====

\section{Modeling Setup} \label{sec:setup}

In this work, we model the interplanetary propagation of the 2015 July 9 CME using the EUropean Heliospheric FORecasting Information Asset \citep[EUHFORIA;][]{pomoell2018} MHD code. Here, we first introduce the inputs and settings necessary to model the solar wind background in Section~\ref{subsec:background}. Then, in Section~\ref{subsec:modelcme} we list the CME input parameters that we derive and set up for running three different EUHFORIA simulations that differ by the configuration---and thus physics---of the modeled ejecta.

\subsection{Solar Wind Background} \label{subsec:background}

The EUHFORIA architecture consists of two modules, namely a coronal domain (1--21.5\,$R_{\odot}$) and a heliospheric one (0.1--2~au)---note that 21.5\,$R_{\odot}$ equals 0.1~au. The coronal domain employs the semi-empirical Wang--Sheeley--Arge \citep[WSA;][]{arge2004} model, which uses synoptic maps of the line-of-sight photospheric magnetic field to compute a background solar wind solution at 21.5\,$R_{\odot}$. Specifically, the WSA model extrapolates the solar magnetic field using the Potential Field Source Surface \citep[PFSS;][]{altschuler1969} model up to 2.6\,$R_{\odot}$ and the Schatten Current Sheet \citep[SCS;][]{schatten1969} model in the range 2.3--21.5\,$R_{\odot}$.

In principle, the ambient solar wind is modeled using a magnetogram from a time before or very close to the CME eruption. For this event, the eruption begins on July 9 around 19:00~UT, and slowly evolves in the low corona to reach 21.5\,$R_{\odot}$ on July 10 after $\sim$08:00~UT as per our forward modeling estimates using GCS (see Figure~\ref{fig:gcs_fits} and Table~\ref{tab:gcs}). Since the filament eruption took place over an extended time period, we consider updating the solar wind conditions around the time the CME reaches the EUHFORIA inner heliospheric boundary of 0.1~au. Therefore, in this work, we employ a magnetogram from July 10 at 00:00~UT to generate the background solar wind. We use a synoptic standard photospheric map from the Global Oscillation Network Group \citep[GONG;][]{harvey1996} of ground-based telescopes.

\subsection{CME Input Parameters} \label{subsec:modelcme}

While keeping the ambient medium described in the previous section fixed, the three EUHFORIA simulation runs that we set up to model the interplanetary propagation of the 2015 July 9 CME differ significantly in the geometric and magnetic configuration of the ejecta. The full list of input parameters used for each run is presented in Table~\ref{tab:euhforia}, and a detailed description of the CME ejecta properties in each run is provided throughout this section.

%%%%% TABLE 2 %%%%%
\begin{table*}[!t]
\centering
\caption{List of the input parameters used to inject each CME in the three different EUHFORIA runs. Latitudes and longitudes are reported in Stonyhurst coordinates. The tilt is measured from the solar west direction and is defined as positive for counterclockwise rotations. Note that for the EUHFORIA+Spheroid run, the 2015 July 9 event is initialized in three parts (see Section~\ref{subsubsec:spheroid} for details).
\label{tab:euhforia}}
\hspace*{-0.18\columnwidth}
\begin{tabular}{lccccc}
\toprule
Model version $\rightarrow$ & \multicolumn{3}{c}{EUHFORIA+Spheroid} & EUHFORIA+Spheromak & EUHFORIA+FRi3D \\
$\downarrow$ Input & Part 1 & Part 2 & Part 3 \\
\midrule
Injection day & 2015-07-10 & 2015-07-10 & 2015-07-10 & 2015-07-10 & 2015-07-10 \\
Time at 21.5\,$R_{\odot}$ ($t_{0}$)& 08:30 & 12:54 & 20:18 & 12:54 & 12:54 \\
Latitude ($\theta$) & $-33^{\circ}$ &  $-35^{\circ}$ & $-35^{\circ}$ & $-35^{\circ}$ & $-38^{\circ}$\\
Longitude ($\phi$) & $-32^{\circ}$ & $-18^{\circ}$ & $40^{\circ}$ & $-18^{\circ}$ & $-25^{\circ}$ \\
Axial tilt ($\gamma$) & $90^{\circ}$ & $22^{\circ}$ & $70^{\circ}$ & $22^{\circ}$ & $10^{\circ}$\\
Nose speed ($V_{0}$) & 560 km$\cdot$s$^{-1}$ & 425 km$\cdot$s$^{-1}$ & 600 km$\cdot$s$^{-1}$ & 425 km$\cdot$s$^{-1}$ & 425 km$\cdot$s$^{-1}$ \\
Semi-major width ($R_\mathrm{maj}$) & $23^{\circ}$ & $43^{\circ}$ & $25^{\circ}$ & --- & $50^{\circ}$ \\
Semi-minor width ($R_\mathrm{min}$) & $18^{\circ}$ & $23^{\circ}$ & $17^{\circ}$ & --- & $26^{\circ}$ \\
Radius ($R_{0}$) & --- & --- & --- & 18.7\,$R_{\odot}$ & --- \\
Toroidal height ($h_{T}$)  & --- & --- & --- & --- & 15.0\,$R_{\odot}$ \\
Mass density ($\rho$) & $10^{-18}$~kg$\cdot$m$^{-3}$ & $10^{-18}$~kg$\cdot$m$^{-3}$ & $10^{-18}$~kg$\cdot$m$^{-3}$ & $10^{-18}$~kg$\cdot$m$^{-3}$ & $10^{-17}$~kg$\cdot$m$^{-3}$ \\
Temperature ($T$) & $8\times10^{5}$~K & $8\times10^{5}$~K & $8\times10^{5}$~K & $8\times10^{5}$~K & $8\times10^{5}$~K \\
Chirality ($\chi$) & --- & --- & --- & +1 & +1 \\
Total flux ($\Phi_{B}$) & --- & --- & --- & $2.0\times10^{13}$~Wb & $2.0\times10^{13}$~Wb\\
Polarity ($\Xi$) & --- & --- & --- & --- & EW \\
Pancaking ($\zeta$)& --- & --- & --- & --- & 0.5 \\
Flattening ($\eta$) & --- & --- & --- & --- & 0.5 \\
Skew ($\psi$) & --- & --- & --- & --- & $30^{\circ}$ \\
Twist ($\tau$) & --- & --- & --- & --- & 1.2 \\
\bottomrule
\end{tabular}
\end{table*}
%%%%%%%%%%%%%%%%%%%

\subsubsection{EUHFORIA+Spheroid} \label{subsubsec:spheroid}

Our first simulation run employs the EUHFORIA+Spheroid \citep{scolini2023} model, which is an extension of the ``original'' EUHFORIA+Cone \citep{pomoell2018} model in that it generalizes the geometry of the CME ejecta front from the ice-cream cone of \citet{fisher1984}---with a circular cross-section---to that of a spheroid---thus, with an elliptical cross-section. In both these EUHFORIA versions, CMEs are modeled as hydrodynamic pulses, with a uniform density distribution across their volume and no internal magnetic field. This means that it is possible to simulate the CME arrival time and speed, but not its magnetic configuration. This approach, albeit simplified, allows us to emulate the highly asymmetric and slow nature of the 2015 July 9 event by injecting the CME into the EUHFORIA heliospheric domain as three separate parts, i.e.\ those identified in the GCS reconstructions showcased in Figure~\ref{fig:gcs_fits}---corresponding to the streamer blowout, the large loop structure, and the three-part cavity (see Section~\ref{subsec:corona}). Therefore, the full CME eruption from an extended filament channel that spans almost the entire solar disk is modeled as simply the superposition of the three separate spheroid eruptions staggered in both longitudinal position and time. This approach is well-suited for the spheroid hydrodynamic pulses because of their lack of an internal magnetic structure, i.e.\ there is no possibility for distinct CME flux systems to interact via e.g.\ collisions and/or reconnection.

The input parameters used for each CME portion inserted in the EUHFORIA+Spheroid simulation run are reported in Table~\ref{tab:euhforia}. The CME injection time ($t_{0}$), latitude ($\theta$), longitude ($\phi$), and tilt ($\gamma$) are taken directly from the GCS fitting results reported in Table~\ref{tab:gcs}. The dimensions of the spheroid, i.e.\ the semi-major ($R_\mathrm{maj}$) and semi-minor ($R_\mathrm{min}$) angular widths of its elliptical cross-section, are derived from the GCS fits by extracting the projection of the 3D structure along the plane perpendicular to the radial direction of the CME nose at its point of maximum breadth (the full details of this procedure are outlined in Appendix~\ref{app:gcs}). The CME injection speed ($V_{0}$) is determined by performing additional GCS fits 1~hour before the times at which each CME part is estimated to have reached 21.5\,$R_{\odot}$, and thus by calculating the velocity needed for the CME front(s) to travel the difference in radial distance (or nose height) between the two sets of reconstructions.

% spheroid: we use the tangent to go from half width to radius

\subsubsection{EUHFORIA+Spheromak} \label{subsubsec:spheromak}

Our second simulation run employs the EUHFORIA+Spheromak \citep{verbeke2019} model, which describes the CME ejecta as a linear force-free spheromak \citep{vandas1997}. Since in this version of EUHFORIA CMEs are magnetized, it is possible to model not only the arrival time of the CME-driven disturbance but also the magnetic structure of the following ejecta. On the other hand, the spherical geometry limits the ability to cover the entire eruption in the case of the large, asymmetric 2015 July 9 event, and the presence of internal fields no longer allows a convenient ``superposition'' of multiple ejecta because of the expected interactions between distinct CME flux systems. Since in this work we are primarily interested in modeling the arrival of the CME at Earth, we resolve to focus our modeling efforts on the portion of the eruption that encompasses Earth's longitude, i.e.\ that corresponding to the loop-like structure shown in Figure~\ref{fig:gcs_fits}(b), or ``Part~2'' in the EUHFORIA+Spheroid simulation.

The input parameters used for the EUHFORIA+Spheromak simulation run are reported in Table~\ref{tab:euhforia}. Since the geometric and kinematic properties of the CME are based on the GCS reconstruction of ``Part~2'' in Table~\ref{tab:gcs}, parameters such as the CME injection time ($t_{0}$), latitude ($\theta$), longitude ($\phi$), tilt ($\gamma$), and nose speed ($V_{0}$) are identical to those in the corresponding portion of CME in the EUHFORIA+Spheroid run. The intrinsic spherical morphology of the spheromak only allows for the set up of a single radius. Given the high latitude of the CME apex as inferred from coronagraph observations, we construct our spheromak around the major axis of the GCS-fitted CME (more information is provided in Appendix~\ref{app:spheromak}). Furthermore, the set of input parameters for the EUHFORIA+Spheromak run include two additional properties that are needed to model the magnetic structure of the embedded flux rope. The first of these is the chirality ($\chi$), set to right-handed based on the analysis presented in Section~\ref{subsec:eruption}. The second of these is the total magnetic flux ($\Phi_{B}$) embedded in the CME, for which we use the reconnection flux of the 2015 July 9 event derived in Paper~I using the flare ribbon masking method of \citet{kazachenko2017}, i.e.\ $2.0{\times}10^{13}$~Wb. We remark that the 2015 July 9 filament eruption, due to its slow and gradual nature, was not characterized by classic flare ribbons signatures, and the reconnection flux was estimated based on EUV emissions more suggestive of post-eruption arcades---the reader can refer to Appendix~B of Paper~I for further details. Finally, we note that the EUHFORIA implementation of the Spheromak CME employed in this work does not take the CME nose speed and total flux as inputs, but rather the radial speed and axial flux. We have chosen to display, nevertheless, $V_{0}$ and $\Phi_{B}$ for uniformity across all the runs shown in Table~\ref{tab:euhforia}---more information on the equations employed to determine instead the Spheromak-required input parameters is outlined in Appendix~\ref{app:spheromak}.

\subsubsection{EUHFORIA+FRi3D} \label{subsubsec:fri3d}

Finally, our third simulation run employs the Flux Rope in 3D \citep[FRi3D;][]{isavnin2016} model to describe the CME ejecta, thus yielding the EUHFORIA+FRi3D \citep{maharana2022} architecture. As in the spheromak case, this ejecta is magnetized, but the overall geometry of the CME---at least at insertion time---is croissant-like (analogous to the GCS morphology) instead of spherical. Furthermore, FRi3D allows for global deformations of the structure due to pancaking \citep[a direct consequence of the radial propagation of CMEs; e.g.,][]{riley2004}, front-flattening \citep[due to the drag exerted by the background solar wind; e.g.,][]{vrsnak2013}, and rotational skew \citep[due to the east--west asymmetry resulting from interaction with a spiral-shaped solar wind; e.g.,][]{luhmann2020}. Despite these additional free parameters, it is not possible to reproduce the full extent of the large 2015 July 9 CME with the FRi3D model, especially considering that the three sub-parts identified in white-light data never appear simultaneously in the coronagraph's field of view (see Figure~\ref{fig:gcs_fits}). Hence, we employ a similar approach as in the spheromak case, i.e.\ we focus on the loop-like structure that encompasses Earth's longitude (``Part~2'' in the EUHFORIA+Spheroid run). For this simulation, our CME input parameters are not derived from the GCS reconstructions in Figure~\ref{fig:gcs_fits}, but from FRi3D's own white-light CME fitting tool---more information is provided in Appendix~\ref{app:fri3d}, and a comparison of the GCS and FRi3D reconstructions is shown in Figure~\ref{fig:gcs+fri3d}.

The input parameters used for the EUHFORIA+FRi3D simulation run are reported in Table~\ref{tab:euhforia}. With respect to the Spheromak run, there are a few additional properties that describe the CME: the toroidal height ($h_{T}$; i.e., the heliocentric distance of the apex of the flux rope axis), the polarity ($\Xi$; i.e., the direction of the flux rope axial field), the deformation parameters described above, i.e., pancaking ($\zeta$), flattening ($\eta$), and skew ($\psi$), as well as the flux rope twist ($\tau$). We set the flux rope polarity from east to west (EW), based on the remote-sensing analysis presented in Section~\ref{subsec:eruption} (see, in particular, Figure~\ref{fig:disc}(b)). For the flux rope twist, which is not possible to determine from solar disc imagery without performing magnetic field extrapolations, we set a value of 1.2 turns based on the statistical analysis of \citet{wang2016}, who found that the most probable estimate for $\tau$ (from leg to leg) is between 0.8 and 1.6 turns. As in the case of the Spheromak run, we use for the total flux $\Phi_{B}$ = $2.0{\times}10^{13}$~Wb based on the results of Paper~I. The remaining parameters (i.e., $h_{T}$, $\zeta$, $\eta$, and $\psi$) are direct outputs of the FRi3D white-light fitting tool. \edit1{$\zeta$ and $\eta$ can assume values in the range [0, 1], with 0.5 corresponding to the ``default'' settings of the flux rope geometry, i.e. without pancaking or front-flattening distortion. $\psi$ acts on the east--west direction only (since it's uniquely due to the spiral nature of the solar wind) and can assume values in the range [$-90^{\circ}$, $90^{\circ}$]. The selected skew of $30^{\circ}$ is certainly higher than what would be employed for the majority of CMEs observed in the corona, but it reflects the high degree of east--west asymmetry experienced by the 2015 July 9 event.}

% ===== MODELLING RESULTS =====

\section{Modeling the CME Propagation} \label{sec:results}

In this section, we present our modeling results in terms of both the large-scale structure of the simulated CMEs through the inner heliosphere and the corresponding (hindcast) predictions at Earth. In Section~\ref{subsec:simresults}, we compare the structure(s) of the CMEs that are injected in EUHFORIA to the structure of the CME in the solar corona that resulted from the modeling efforts (of the filament's eruption and early evolution) presented in Paper~I. In Section~\ref{subsec:simearth}, we compare the synthetic time series at Earth from the three EUHFORIA runs---alongside the so-called ambient run of the steady-state background---with in-situ observations from the Wind spacecraft.

\subsection{Large-Scale Structure of the CME} \label{subsec:simresults}

%%%%% FIGURE 4 %%%%%
\begin{sidewaysfigure*}[p!]
\centering
\vskip -3.9in
	\includegraphics[width=\linewidth]{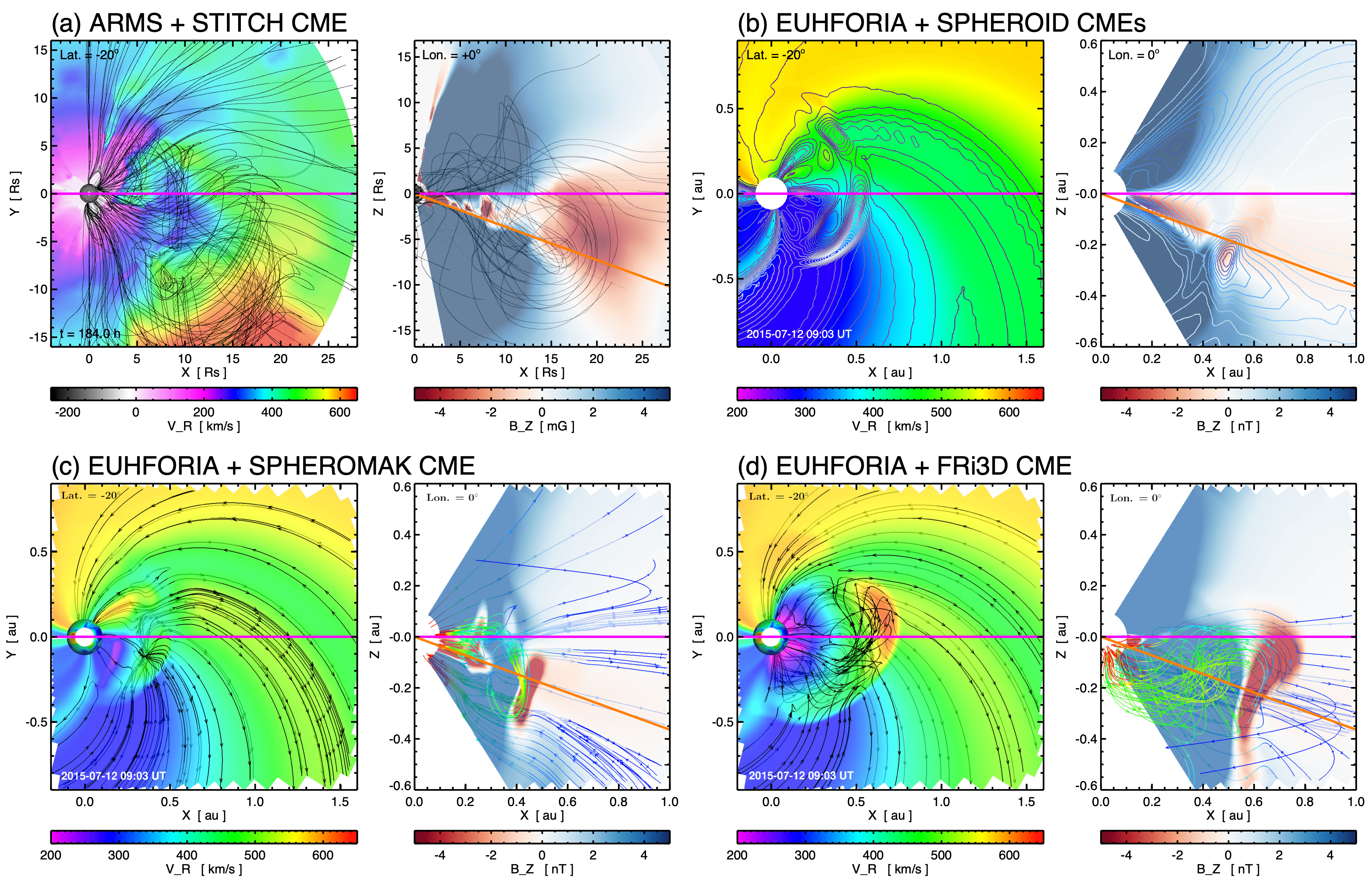}
	\caption{Overall modeling results of the 2015 July 9 CME. The figure shows the three heliospheric simulations performed with EUHFORIA compared with each other and with the Paper~I coronal simulation modeled with ARMS+STITCH. (a) ARMS+STITCH simulation. (b) EUHFORIA+Spheroid simulation. (c) EUHFORIA+Spheromak simulation. (d) EUHFORIA+FRi3D simulation. Each panel shows (left) the $-20^{\circ}$ latitude plane and (right) the $0^{\circ}$ longitude plane in Stonyhurst coordinates. The latitudinal plane shows the radial velocity $V_r$ and the longitudinal plane shows the $B_z$ component of the magnetic field. Representative magnetic field lines are shown in the simulations with magnetized CME ejecta and the black contours in panel (b) highlight the merged density enhancement of the non-magnetized spheroid CMEs.}
	\label{fig:allresults}
\end{sidewaysfigure*}
%%%%%%%%%%%%%%%%%%%

First of all, our main focus is to evaluate the large-scale structure and morphology of the modeled event in each of the three EUHFORIA simulation runs, especially to assess similarities and differences. Additionally, we are interested in comparing the EUHFORIA heliospheric results with the coronal ones obtained in Paper~I. The simulation presented and described in Paper~I was performed using the Adaptively Refined MHD Solver \citep[ARMS;][]{devore2008} code, which solves the equations of ideal MHD on a 3D spherical grid with solution-adaptive mesh refinement implemented via the PARAMESH \citep{macneice2000} toolkit. In Paper~I, the pre-eruption corona was modeled with a low-order spherical harmonic expansion based on the GONG synoptic map for Carrington Rotation 2165 with an isothermal \citet{parker1958} solar wind for $T=1.4$~MK. The extended filament channel was energized with the STatistical InjecTion of Condensed Helicity \citep[STITCH;][]{dahlin2022} procedure to introduce sheared horizontal flux along the PIL. The filament eruption begins from the eastern limb and progresses towards the western limb, resulting in a skewed, asymmetric CME flux rope structure, propagating approximately $30^\circ$ south of the ecliptic plane but expanding significantly in latitude, in agreement with remote-sensing observations of the solar disk and corona.  

Figure~\ref{fig:allresults} presents an overview of the 2015 July 9 CME modeled with the four different simulations in the planes of (left) $-20^{\circ}$ latitude and (right) $0^{\circ}$ longitude (in Stonyhurst coordinates). Within each run, the left panel shows the radial solar wind velocity ($V_r$) and the right panels show the north--south component of the magnetic field ($B_z$). The choice to display an inclined, constant-latitude plane rather than the equatorial (or ecliptic) one is motivated by the fact that, as shown in Paper~I and in the remote-sensing analysis part of Section~\ref{sec:observations}, the bulk of the CME propagated toward the south. A plane inclined by $20^{\circ}$ to the equator allows us to more readily compare the full longitudinal extent of the modeled ejecta, while keeping our viewpoint reasonably close to Earth's latitude. 

From the $V_r$ panels, it is possible to note that there are significant differences arising between the four runs. In the ARMS+STITCH simulation (Figure~\ref{fig:allresults}(a)), there is a clear east--west asymmetry that reflects the observed evolution of the eruption in remote-sensing data, i.e.\ with the east limb portion of the extended filament erupting first, and the rest of the energized filament channel field structure following suit via the well-known zipper effect associated with reconnection above extended PILs \citep[e.g.,][]{priest2017}. This feature of the coronal simulation, i.e.\ the eastern flank/leg of the erupting structure leading with the western one trailing, is not maintained in the heliospheric domain of any of the three EUHFORIA runs (Figure~\ref{fig:allresults}(b--d)). In fact, the western leg is now the leading portion of the ejecta, with the degree of asymmetry depending on the specific run. This holds true even in the case of the EUHFORIA+FRi3D simulation (Figure~\ref{fig:allresults}(d)), in which the CME was injected with an initial morphology that more closely resembled that of the eruption modeled in the corona with ARMS, i.e.\ with a skew of its toroidal axis toward eastern heliolongitudes. In all cases, we can attribute this apparent ``counter-skewing'' in interplanetary space to a fast solar wind stream rooted on the western hemisphere of the Sun---and visible in all the EUHFORIA runs, since they all used the same ambient medium as input. As a consequence, the western flank of the CME propagates faster than the eastern one, thus becoming the leading portion of the overall structure. We also note that interaction with the faster solar wind flow leads to greater acceleration of the ejecta in the EUHFORIA+FRi3D run than in the remaining two. This is likely due to the specific geometry and magnetic description of each of the CME ejecta, which may in turn yield different interaction outcomes. For example, in the EUHFORIA+Spheroid run (Figure~\ref{fig:allresults}(b)), the lack of a magnetized ejecta is not expected to result in a realistic interaction process \citep[see also][]{palmerio2022a}, but rather in a CME-like disturbance propagating more or less straightforwardly through a background wind. In the EUHFORIA+Spheromak run (Figure~\ref{fig:allresults}(c)), the lack of CME legs may represent less of an ``obstacle'' to the ambient solar wind, which in turn may be able to flow about the ejecta with minimal interaction of the two flux systems.

From the $B_z$ panels, it is obvious (as expected) that the CME propagates mostly south of the Sun--Earth line. Additionally, the ARMS and EUHFORIA runs---excluding the Spheroid one, which does not include CME internal fields---all display a leading (trailing) negative (positive) $B_{z}$, in agreement with the SWN flux rope type inferred from remote-sensing observations (see Section~\ref{subsec:eruption}) and, thus, indicating that no dramatic rotations of the ejecta have taken place in the simulation(s). Overall, the EUHFORIA runs display a flatter front than the ARMS one, which is to be expected since the drag exerted by the background solar wind builds up as a CME travels away from the Sun, i.e., its effects are more noticeable in interplanetary space than in the corona. Finally, it appears that the CME in the FRi3D run intercepts the Sun--Earth line with a larger portion than in the remaining two EUHFORIA runs, i.e., its overall trajectory is more comparable to that of the ARMS CME. Again, this may be due to more realistic CME interactions with the ambient wind resulting from the initial configuration of the ejecta magnetic fields.

\subsection{Modeled Time Series at Earth} \label{subsec:simearth}

\begin{figure*}[!t]
\center
	\includegraphics[width=.99\linewidth]{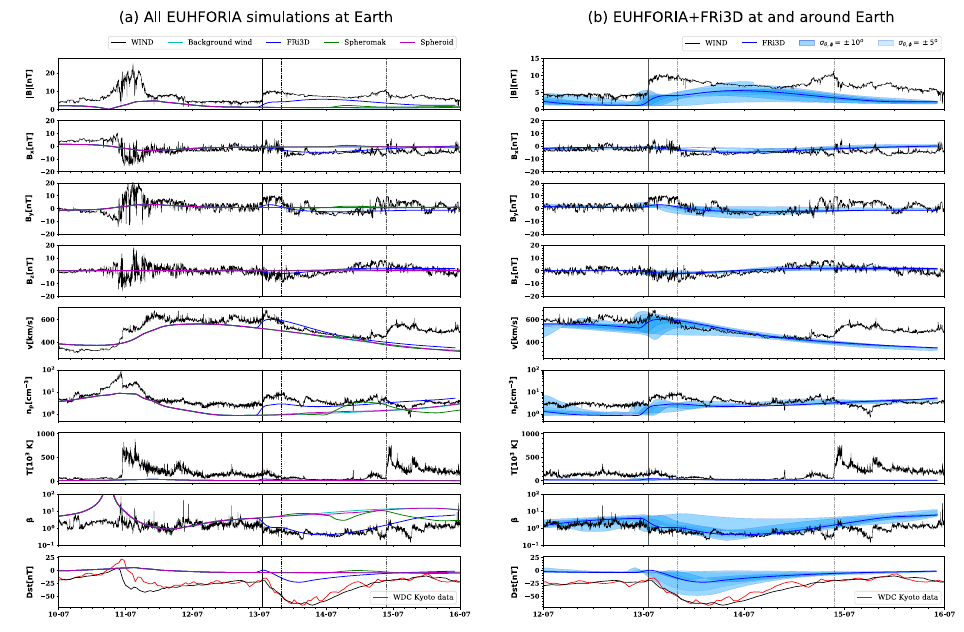}
	\caption{Simulated magnetic field and plasma time series for (a) all EUHFORIA simulations and (b) the EUHFORIA+FRi3D simulation, shown against the corresponding measurements taken by the Wind spacecraft. The panels show, from top to bottom: Magnetic field magnitude, magnetic field components in Cartesian coordinates, solar wind bulk speed, proton density, proton temperature, plasma beta, \edit1{and Dst index \citep[where the red profile features WDC Kyoto data and the black profile displays the modeled Dst based on Wind in-situ measurements using the algorithm of][]{obrien2000}}. The solid vertical lines mark the arrival of the shock-like disturbance, while the dashed vertical lines mark the leading \edit1{and trailing edges} of the ICME ejecta (i.e., the beginning \edit1{and end} of the gray shaded area in Figure~\ref{fig:insitu}). The results in (a) show the simulated time series at Earth only, while the results in (b) show simulation outputs at Earth as well as in the ${\pm}5^{\circ}$ and ${\pm}10^{\circ}$ latitudinal and longitudinal offsets around Earth.}
	\label{fig:euhforia_sims}
\end{figure*}

After evaluating the global structure of the 2015 July 9 CME modeled with EUHFORIA, we analyze the corresponding time series at Earth's location and compare them with in-situ measurements taken by the Wind spacecraft (and described in detail in Section~\ref{subsec:earth}). The collective of in-situ profiles, together with the ambient run (of the background wind with no CMEs), is shown in Figure~\ref{fig:euhforia_sims}(a). First of all, we note that the arrival time and speed of the fast solar wind stream preceding the CME---and that leads to the interaction discussed in Section~\ref{subsec:simresults}---are reproduced fairly well by EUHFORIA. The compression of solar wind material at the stream interaction region is largely underestimated, which is on the other hand a well-known issue of heliospheric MHD models \citep[e.g.,][]{pahud2012, riley2012}.

%%%%% FIGURE 5 %%%%%
%%%%%%%%%%%%%%%%%%%

Regarding the arrival of the CME itself, we note clear differences among the three EUHFORIA runs. The Spheroid simulation, in particular, is hardly distinguishable from the ambient one---the two curves begin separating around 16:00~UT on July 13 (visible especially in the speed, density, and beta profiles), but their offset is so modest that it can be considered negligible. In practice, this event would have been effectively considered a ``miss'' at Earth in terms of space weather predictions using the spheroid model. The Spheromak simulation, on the other hand, features a somewhat clearer impact around 02:00~UT on July 14, i.e.\ approximately one day later than observed (the CME-related disturbance was determined to have commenced around 01:00~UT on July 13 in Section~\ref{subsec:earth}). Furthermore, the CME-induced enhancements in the magnetic field and plasma parameters are substantially less prominent than observed---in practical terms, this prediction would have likely been considered an ``almost-miss'' given its weak impact. Finally, the FRi3D simulation displays an arrival time at Earth that is just about 2~hr earlier than detected by Wind, i.e.\ around 23:00~UT on July 12. Both the plasma structure and magnetic field configuration of the ICME ejecta are generally well captured by the model, but the magnetic field magnitude, as well as the intensity of the $B_{z}$ component, are moderately underestimated. One possible cause for this is that the trailing portion of the fast solar wind stream in which the CME is embedded does not appear in the EUHFORIA simulation(s), leading to a lack of compression at the back of the ejecta. \citet{hinterreiter2019}, in a study aimed at assessing the performance of EUHFORIA in modeling the ambient background, found that coronal hole areas are generally underestimated, often yielding narrower flows of fast wind than observed in in-situ data.

A zoomed-in view of the CME arrival in the FRi3D simulation is shown in Figure~\ref{fig:euhforia_sims}(b) alongside results at virtual spacecraft surrounding Earth with offsets of ${\pm}5^{\circ}$ and ${\pm}10^{\circ}$ in both latitude and longitude. As discussed in \citet{scolini2019}, considering outputs around the specific location of interest is an adequate way to account for the typical uncertainties associated with estimates of CME size and/or trajectory based on remote-sensing imagery \citep[which become even more prominent when employing a single viewpoint; see][]{verbeke2022}. Across the whole range of virtual spacecraft, the CME arrival times are clustered ${\pm}8$~hr around the observed one \citep[well within the known typical uncertainties; e.g.,][]{riley2018, wold2018}. The largest spreads in predictions of the ICME ejecta are found in the magnetic field magnitude and $B_{y}$ component, with values in the range [2, 8]~nT and [$-$8, 2]~nT, respectively. On the other hand, we find no significant differences in the $B_{z}$ component, i.e., no synthetic time series reaches the minimum observed values of $-$9~nT. Thus, although an encounter with the northern flank of the ICME ejecta as well as its overall magnetic configuration are correctly predicted, the intensity of the (moderate) geomagnetic storm that ensued would have been somewhat underestimated.

\edit1{To quantify the geoeffectiveness resulting from each of the EUHFORIA predictions, we show in the bottom panels of Figure~\ref{fig:euhforia_sims} the Dst index, both from WDC Kyoto data and using the algorithm of \citet{obrien2000} with Wind measurements as input (both time series feature an equal value of Dst$_\mathrm{min} = -68$~nT associated with the 2015 July 9 CME). Employing the same algorithm to calculate the Dst from EUHFORIA (Figure~\ref{fig:euhforia_sims}(a)), we note that the Spheroid and Spheromak runs predict no geomagnetic response, while the FRi3D simulation at Earth features Dst$_\mathrm{min} = -25$~nT, suggesting a fairly minor storm. When considering the cluster of virtual spacecraft surrounding Earth (Figure~\ref{fig:euhforia_sims}(b)), we find that all the Dst$_\mathrm{min}$ predictions fall between 0~nT and $-50$~nT, corresponding to a none-to-minor storm response and confirming that the weaker-than-observed modeled $B_{z}$ magnitude (likely due to the lack of compression of the CME from the following fast solar wind stream) leads to an underestimation of the resulting geomagnetic effects.} 

% ===== DISCUSSION =====

\section{Discussion} \label{sec:discussion}

From the results presented in Section~\ref{sec:results}, it is evident that the EUHFORIA+FRi3D run performed significantly better than the remaining two in terms of capturing the arrival and magnetic field configuration of the 2015 July 9 CME. Nevertheless, it is important to scrutinize these findings in the context of extreme events in terms of their spatial and temporal extent. The CME analyzed here erupted over a prolonged amount of time and covered a wide range of heliolongitudes (basically encompassing almost the full Earth-facing solar disk). Therefore, a number of assumptions and approximations had to be made in order to accommodate the heliospheric modeling of such a peculiar event. To emulate the highly asymmetric and slow coronal evolution of the eruption, we divided the whole CME structure into three parts (based on its appearance in white light) in the simulation that did not include internal ejecta magnetic fields (EUHFORIA+Spheroid run, see Section~\ref{subsubsec:spheroid}). For the simulations that featured magnetized ejecta, on the other hand, we resolved to select a ``portion of interest'' to model, as to avoid spurious interactions arising from simulating three separate CME flux systems. The first of them (EUHFORIA+Spheromak run, see Section~\ref{subsubsec:spheromak}) simplified the Earth-directed portion of the event as a magnetized sphere, while the second (EUHFORIA+FRi3D run, see Section~\ref{subsubsec:fri3d}) approximated the structure as an asymmetric croissant that was however significantly less extended in longitude than observed in coronagraph imagery.

The 2015 July 9 CME was encountered at Earth only through its northern edge, but the impact was still relatively close to its nose (under the assumption that the axis of the embedded flux rope was approximately east--west oriented). Had the encounter taken place closer to the apex in latitude, but farther from it in longitude, then the Spheromak run may have yielded more accurate results. This is because the CME would have likely encountered Earth through its leg in the FRi3D run even though the full event extended beyond the boundary of the FRi3D flux rope geometry, whereas the Spheromak configuration lacks any CME leg features \citep[see also][]{maharana2023}. Hence, the simplified Spheromak magnetic structure may be a more convenient modeling strategy when simulating only a portion of a particular CME, depending on the in-situ observer's location. On the other hand, had we been interested in predicting the CME arrival at several observers in the inner heliosphere spread across broad longitudes, then the Spheroid ejecta would have likely been a more fitting option. In fact, the superposition of multiple hydrodynamic pulses launched with different trajectories and at different times results in a more asymmetric and wider ``global'' ejecta, which is more suitable for comparisons with multi-point measurements. Thus, the Spheroid (or Cone) ejecta seems to be currently the most feasible tool to model extended and asymmetric CMEs as a whole. However, the advantages of this approach may be somewhat limited to periods of relatively simple ambient conditions, e.g.\ without significant CME interactions with the structured solar wind that would require a more complete ejecta description to model accurately. Therefore, we caution against extrapolating conclusions that are contingent on the particular event analyzed in this work and remark on the importance of carefully selecting the modeling strategy that best fits the specific research (or operational) goal(s).

Our interest in the 2015 July 9 CME was driven in particular by the moderate problem geomagnetic storm that followed upon its impact at Earth. While this work showcases some of the difficulties associated with this particular class of events, the CME studied here featured some specific characteristics that made it ``problematic'': It was exceptionally extended as well as highly asymmetric (and, thus, difficult to model in its entirety), it was slow, and had no significant Earth-directed component (at least according to estimates from remote-sensing imagery). Furthermore, its related space weather effects were likely intensified by the fact that it traveled through a flow of high-speed solar wind---that CME interaction with fast streams can lead to enhanced geoeffectiveness has been shown both via observational \citep[e.g.,][]{nitta2021, palmerio2022b} and modeling \citep[e.g.,][]{kay2022} studies. Our work demonstrated two viable ways as to how to model cases in which the ``problematic'' status can be attributed to an extended and/or asymmetric structure of the CME, i.e., either via modeling only the Earth-directed (or observer-of-interest-directed) component or via constructing the whole CME from separate parts. This also raises interesting questions regarding the physics and coherence of more extended CMEs. For example, are CMEs truly composed of distinct sections or simply appear complex in coronagraph images due to local distortions and/or projection effects, and to what extent do they maintain their integrity while they propagate in the structured solar wind? Multi-spacecraft heliospheric observations over broad longitudes and short radial distances \citep[e.g.,][]{lugaz2022} may provide further insight into these issues.

Nevertheless, problem geomagnetic storms can feature different characteristics of their solar counterparts that grant their ``problematicness,'' including the weakness or total absence of remote-sensing signatures \citep[e.g.,][]{nitta2017, palmerio2021b}. While attempts at modeling such events have been performed \citep[e.g.,][]{lynch2016b, palmerio2021c}, these CMEs are mostly analyzed in hindcast mode, as they often escape the eyes in real-time applications. In the case of the 2015 July CME investigated here, on the other hand, the problematic nature of the event in terms of real-time predictions was given by the fact that the CME appeared as three separate eruptions in white-light imagery, and a detailed analysis of the corresponding solar disk observations was necessary to gain a deeper insight into its onset and early evolution. In the future, improvements in resolution and sensitivity for EUV and white-light imaging, as well as the availability of remote-sensing observers from multiple viewpoints \citep[see, e.g.,][]{gibson2018a, howard2023, palmerio2023} will significantly advance our ability to identify and include these different types of problematic CMEs in space weather forecasting applications.

% ===== SUMMARY AND CONCLUSIONS =====

\section{Summary and Conclusions} \label{sec:conclusions}

In this work, we have simulated the interplanetary propagation and arrival at Earth of the large, slow, and highly asymmetric CME that erupted on 2015 July 9. Starting from estimates of the CME geometry and kinematics based on white-light coronagraph observations, we have adopted three different approaches to represent the ejecta within the EUHFORIA model. In the first (EUHFORIA+Spheroid, lacking an internal ejecta magnetic field) we modeled the event in three separate parts to emulate the wide spread in longitude and slow temporal evolution of the CME. In the remaining two (EUHFORIA+Spheromak and EUHFORIA+FRi3D, both featuring magnetized ejecta but with different morphologies) we selected the Earth-directed portion of the exceptionally large CME for heliospheric modeling of its propagation. We found that the FRi3D simulation run yielded the best results at Earth in terms of CME arrival time and magnetic configuration upon impact, but remarked that the remaining two ejecta descriptions may represent better modeling choices in different situations, especially in the case of forecasting the properties of problem geomagnetic storms.

One major challenge in interplanetary CME modeling---and especially evident in the case of the 2015 July 9 event---is represented by adapting complex CME morphologies to the simplified flux rope models that are implemented in existing heliospheric simulations. While flux rope models of increasingly higher complexity are being developed \citep[e.g.,][]{weiss2022, nieveschinchilla2023}, their inclusion within global MHD simulations is still a considerable challenge from a computational standpoint. Alternative, but more complex approaches, include full Sun-to-heliosphere simulations \citep[e.g.,][]{jin2017, torok2018}---in our case, this would translate into propagating the 2015 July 9 CME as modeled with ARMS in Paper~I seamlessly from the coronal domain into the heliosphere. However, such simulation runs are computationally expensive and, thus, less suitable for real-time space weather predictions at least for the time being. The approach undertaken in this work represents a practical way to circumnavigate these current limitations and to provide, in the meanwhile, a strategy toward modeling more problematic events.

% ===== ACKNOWLEDGEMENTS =====

\section*{Acknowledgments}
%\begin{acknowledgments}
E.P.\ acknowledges support from NASA's HTMS (grant no.\ 80NSSC20K1274) and LWS-SC (grant no.\ 80NSSC22K0893) programs as well as NSF's PREEVENTS (grant no.\ ICER-1854790) program.
A.M.\ acknowledges support from projects C14/19/089 (C1 project Internal Funds KU Leuven), G.0025.23N (WEAVE FWO-Vlaanderen), SIDC Data Exploitation (ESA Prodex-12), and Belspo project B2/191/P1/SWiM.
B.J.L.\ acknowledges NSF AGS-2147399, NASA 80NSSC22K0674, and NASA 80NSSC21K1325.
C.S.\ acknowledges support from the Research Foundation -- Flanders (FWO strategic base PhD fellowship 1S42817N) and NASA grants 80NSSC19K0914, 80NSSC20K0197, and 80NSSC20K0700.
S.W.G.\ was supported by Academy of Finland Fellowship grants 338486 and 346612 (INERTUM), and Project grant 310445 (SMASH).
S.W.G., J.P., and E.K.J.K.\ acknowledge the European Research Council (ERC) under the European Union's Horizon 2020 Research and Innovation Programme Project SolMAG (grant agreement no.\ 724391) and the Finnish Centre of Excellence in Research of Sustainable Space (Academy of Finland grants no.\ 312390 and no.\ 336807).
EUHFORIA is developed as a joint effort between the University of Helsinki and KU Leuven. Simulations were carried out at KU Leuven and the Flemish Supercomputer Center (VSC), funded by the Hercules foundation and the Flemish Government, Department of Economy, Science and Innovation (EWI).
The CACTus CME catalog (\url{http://sidc.oma.be/cactus/}) is generated and maintained by the SIDC at the Royal Observatory of Belgium. The SOHO/LASCO CME catalog (\url{https://cdaw.gsfc.nasa.gov/CME_list/}) is generated and maintained at the CDAW Data Center by NASA and The Catholic University of America in cooperation with the Naval Research Laboratory.
We thank the WDC for Geomagnetism, Kyoto (\url{http://wdc.kugi.kyoto-u.ac.jp}), and the geomagnetic observatories for their cooperation to make the final Dst indices available.
E.P.\ thanks the Croom team for fruitful discussions and collaborations.
We also thank Dr.\ Daria Shukhobodskaia (Royal Observatory of Belgium, Brussels) for developing the user-friendly GUI application used in this work to fit the FRi3D model to white-light images.
%\end{acknowledgments}

\facilities{ACE (SWICS); BBSO (H$\alpha$); GONG (Magnetogram); SOHO (LASCO); SDO (AIA, HMI); Wind (MFI, SWE)}
\software{ARMS \citep{devore2008}; AstroPy \citep{astropy2022}; EUHFORIA \citep{pomoell2018}; SolarSoft \citep{freeland1998}; SunPy \citep{sunpy2020}}

%\clearpage

% ===== APPENDIX =====
\appendix

\section{Derivation of the CME Angular Widths from GCS Fits} \label{app:gcs}

\begin{figure}[!t]
\center
	\includegraphics[width=0.75\linewidth]{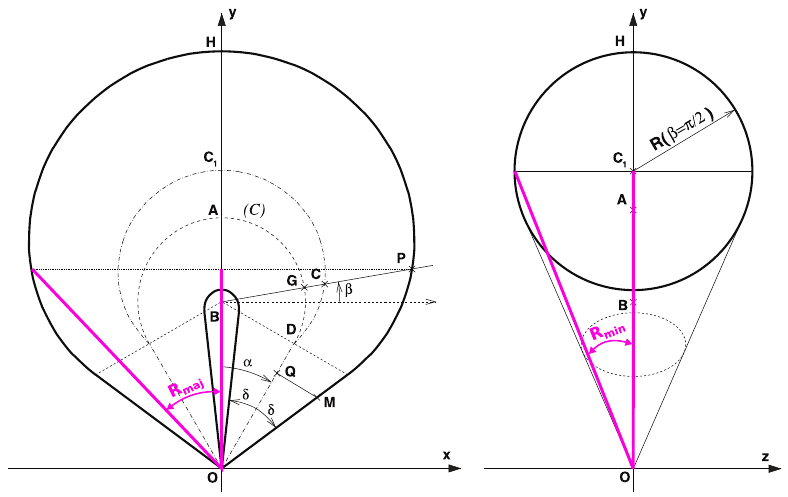}
	\caption{Overview of the GCS geometry in the (left) face-on and (right) edge-on projections, adapted from \citet{thernisien2011}. The half-angular face-on ($R_\mathrm{maj}$) and edge-on ($R_\mathrm{min}$) widths of the croissant are highlighted in magenta.}
	\label{fig:gcs_model}
\end{figure}

In the following, we refer to the formulation of \citet{thernisien2011}, who provides an overview of the GCS croissant geometry and all its parameters, summarized in Figure~\ref{fig:gcs_model}. Given the croissant-like geometry of the GCS model, it follows that the projection of the 3D wireframe on the plane perpendicular to the radial direction of the CME nose at its maximum extent corresponds to an ellipse of semi-major axis $R_\mathrm{maj}$ and semi-minor axis $R_\mathrm{min}$ (highlighted in Figure~\ref{fig:gcs_model}), corresponding to the face-on and edge-on angular widths, respectively.

%%%%% FIGURE A1 %%%%%
%%%%%%%%%%%%%%%%%%%

Considering that the GCS aspect ratio parameter is defined as $\kappa = \sin\delta$, one obtains the half-angular width of the edge-on cross section ($R_\mathrm{min}$) from Equations~(28) and (29) in \citet{thernisien2011}, as well as the relation $\tan{R_\mathrm{min}} = R(\beta=\pi/2)/OC_1$ (right panel of Figure~\ref{fig:gcs_model}), which gives
\begin{equation}
R_\mathrm{min} = \arctan{\kappa}\, .
\label{eqn:weo}
\end{equation}

\citet{thernisien2011} outlined the procedure for the numerical calculation of the face-on half-width $R_\mathrm{maj}$ by noting that it corresponds to the maximum of the $x$-coordinate of $\overrightarrow{BP}$. Thus, $R_\mathrm{maj}$ is estimated by finding the value of the angle $\beta$ such that $\partial (\overrightarrow{BP}\cdot \hat{x})/\partial \beta = 0$.
Given that $\overrightarrow{BP}\cdot \hat{x} = \left( X_0 + R \right) \cos{\beta}$ using the \citet{thernisien2011} definitions

\begin{equation}
X_0 = \frac{\rho + b \kappa^2 \sin{\beta}}{1-\kappa^2} \;\;\;\; \text{and} \;\;\;\; R = \left[ X_0^2 + \frac{b^2\kappa^2 - \rho^2}{1-\kappa^2} \right]^{1/2} \; ,
\label{eq:a2}
\end{equation} 

\noindent setting this derivative to zero yields

\begin{eqnarray}
\label{eq:deriv}
\frac{\partial}{\partial \beta} \left( \overrightarrow{BP}\cdot \hat{x} \right) &=& \frac{\partial}{\partial \beta} f(\beta) \nonumber \\
&=& \frac{\partial}{\partial \beta} 
\left[ \left( \frac{\rho + b \kappa^2 \sin{\beta}}{1-\kappa^2} \right) \cos{\beta} + \left( \left( \frac{\rho + b \kappa^2 \sin{\beta}}{1-\kappa^2} \right)^2 + \frac{b^2\kappa^2 - \rho^2}{1-\kappa^2} \right)^{1/2} \cos{\beta} \right] = 0 \; .
\label{eq:a3}
\end{eqnarray}

\begin{figure}[!t]
\center
	\includegraphics[width=0.9\linewidth]{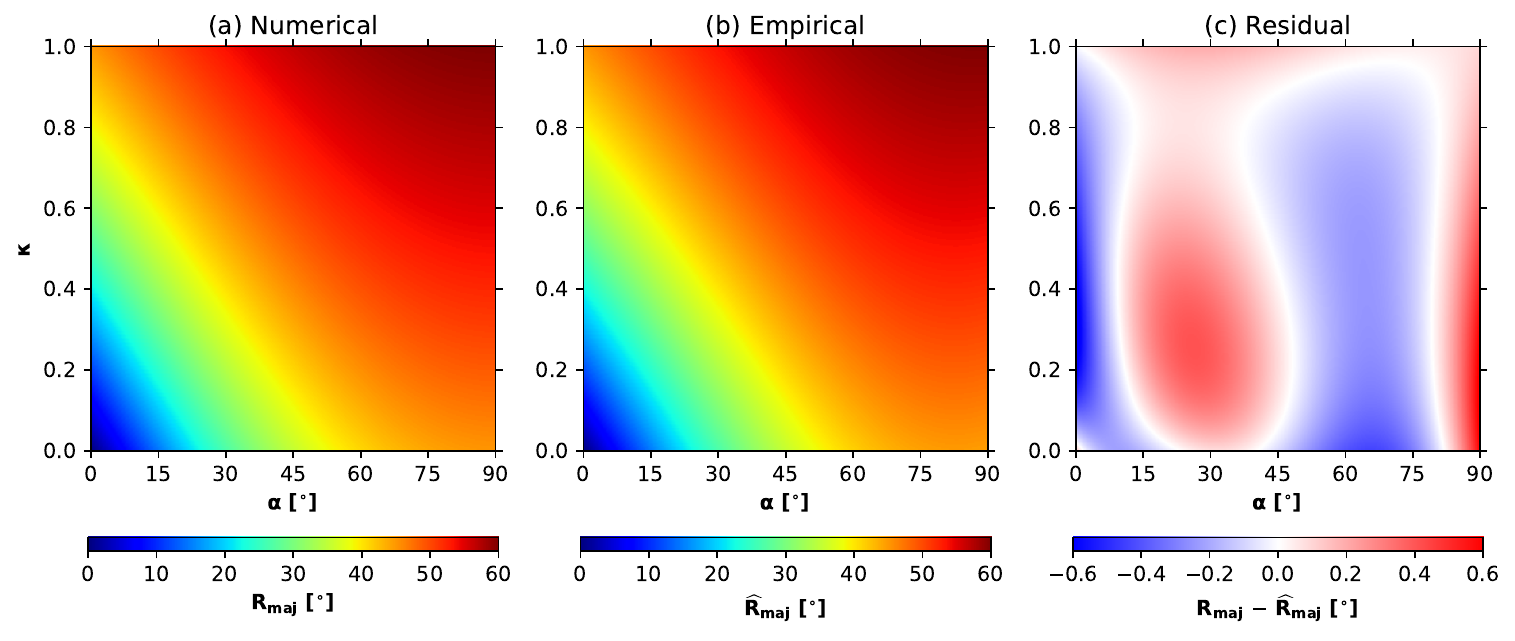}
	\caption{Dependence of the GCS face-on width $R_\mathrm{maj}$ on model parameters $\alpha$ and $\kappa$. (a) $R_\mathrm{maj}$ computed numerically via Equation~\ref{eq:a4} using the procedure outlined in \citet{thernisien2011}. (b) The empirical fit $\widehat{R}_\mathrm{maj}$ given by Equation~\ref{eqn:wfo}. (c) The residual $(R_\mathrm{maj}-\widehat{R}_\mathrm{maj})$ between the numerical (a) and empirical (b) formulations.}
	\label{fig:wfo}
\end{figure}

\noindent We note that with the substitution of the additional GCS geometric parameter definitions $b = h / \cos{\alpha}$ and $\rho = h \tan{\alpha}$, the expression for the derivative is now solely a function of the parameters $(\alpha, \kappa, h)$. Since $h$ describes the full height of the cone under self-similar expansion ($OD$ in Figure~\ref{fig:gcs_model}), it does not affect the angular dependence. We have solved Equation~\ref{eq:deriv} numerically
%using {\tt fsolve} in {\tt matplotlib} for Python 
to obtain values of $\beta_0$ such that $\partial f(\beta_0)/\partial \beta = 0$ for a $300 \times 300$ input array of $(\alpha, \kappa)$ with $\alpha \in [0^{\circ}, 90^{\circ}]$ and $\kappa \in [0,1]$. This yields a 2D distribution of $R_\mathrm{maj}(\alpha,\kappa)$ given by

\begin{equation}
R_\mathrm{maj}(\alpha,\kappa) = \arcsin{\left[ \frac{ | \overrightarrow{BP} | \cos\beta_0} { \sqrt{b^2 + |\overrightarrow{BP}|^2 + 2 b | \overrightarrow{BP} | \sin\beta_0 } } \right]} \; ,
\label{eq:a4}
\end{equation}

\noindent where $| \overrightarrow{BP} | = X_0(\beta_0) + R(\beta_0)$. Figure~\ref{fig:wfo}(a) shows the numerical $R_\mathrm{maj}(\alpha, \kappa)$ results. Since this procedure can be somewhat cumbersome, we have also constructed an empirical approximation,  $\widehat{R}_\mathrm{maj}(\alpha, \kappa)$, as a 2nd-order, 2D polynominal fit to the numerical $R_\mathrm{maj}$, 

\begin{align}
\widehat{R}_\mathrm{maj}(\alpha, \kappa) = \sum_{i,j=0}^{2} c_{ij}\,\alpha^{i}\,\kappa^{j} &=
\begin{bmatrix}
 \alpha^{0}  & \alpha^{1} & \alpha^{2}
\end{bmatrix}
\begin{bmatrix}
%  -9.17{\times}10^{-2} & +1.08{\times}10^{0} & -6.53{\times}10^{-3} \\
%  +6.38{\times}10^{1} & -1.04{\times}10^{0} & +6.13{\times}10^{-3} \\
%  -1.87{\times}10^{1} & +2.98{\times}10^{-1} & -1.52{\times}10^{-3}
    -9.17{\rm E}{-02} & +1.08{\rm E}{+00} & -6.53{\rm E}{-03} \\
    +6.38{\rm E}{+01} & -1.04{\rm E}{+00} & +6.13{\rm E}{-03} \\
    -1.87{\rm E}{+01} & +2.98{\rm E}{-01} & -1.52{\rm E}{-03}
\end{bmatrix}
\begin{bmatrix}
  \kappa^{0}  \\
  \kappa^{1}  \\
  \kappa^{2} \\
\end{bmatrix} \; .
\label{eqn:wfo}
\end{align}

%%%%% FIGURE A2 %%%%%
%%%%%%%%%%%%%%%%%%%

\noindent Figure~\ref{fig:wfo}(b) shows this empirical estimate $\widehat{R}_\mathrm{maj}(\alpha, \kappa)$, and Figure~\ref{fig:wfo}(c) plots the residual $R_\mathrm{maj} - \widehat{R}_\mathrm{maj}$. The magnitude of the residual is always ${\lesssim}0.6^{\circ}$, which is significantly less than the inherent uncertainty associated with the GCS geometry parameters due to both the model's sensitivity \citep{thernisien2009} and the user's subjectivity \citep{verbeke2022}.

\section{Derivation of Spheromak-Specific Input Parameters} \label{app:spheromak}

The GCS model usually features an elongated (croissant-like) morphology, unless $\alpha=0$, for which $R_\mathrm{maj} = R_\mathrm{min}$ and the so-called cone model is retrieved. When adapting GCS parameters to the spheromak morphology, a single radius has to be derived, usually from the half-width $\alpha$ and/or the aspect ratio $\kappa$, for cases in which $\alpha\neq0$. In this work, we employ for the spheromak run the ``maximum'' (face-on) radius derived from GCS parameters as defined by \citet{asvestari2021},

\begin{equation}
R_{0} = 21.5\,R_{\odot} \sin \left( \frac{ \omega_\mathrm{FO}}{2}\right) \; ,
\label{eq:spheromakradius}
\end{equation}

\noindent where $\omega_\mathrm{FO} = 2 (\alpha + \delta) = 2 (\alpha + \arcsin{\kappa})$ indicates the full face-on width of the croissant (see Figure~\ref{fig:gcs_model}).

The spheromak radial speed ($V_\mathrm{rad}$) is derived from the CME nose speed ($V_{0}$) using the formula in \citet{scolini2019}, who derived that the total (nose) speed of a CME front can be decomposed into its radial speed (i.e., the speed of the CME center) and its expansion speed (i.e., the rate of growth of its cross-section). In GCS quantities, this converts to
\begin{equation}
V_{0} = V_\mathrm{rad} + V_\mathrm{exp} = \frac{1}{1+\kappa} \diff{H}{t} + \frac{\kappa}{1+\kappa} \diff{H}{t} \; ,
\label{eq:spheromakspeed}
\end{equation}
where $\kappa$ is again the GCS aspect ratio, and $H$ is the height of the CME nose ($OH$ in Figure~\ref{fig:gcs_model}).

Another parameter that has to be adapted to the spheromak implementation of EUHFORIA is the axial (or toroidal) magnetic flux ($\Phi_{T}$) of the CME. To do so, we start from the total reconnection flux ($\Phi_{B}$) estimated in Paper~I and use the formulation of \citet{scolini2019} under the assumption that all the reconnected flux goes into the poloidal magnetic flux of the erupted flux rope \citep{qiu2007} to obtain the toroidal flux as:

\begin{equation}
\Phi_{T} = \frac{2B_{0}}{a^{2}} \left[ -\sin(x_{01}) + \int_{0}^{x_{01}} \frac{\sin x}{x} dx \right] \; ,
\label{eq:spheromakflux}
\end{equation}

where $\alpha$ is the force-free parameter and $x_{01} = \alpha R_{0} = 4.4934$ is the first zero of the Bessel function $J_{1}$. $B_{0}$ is  the axial field strength of the spheromak flux rope, defined as

\begin{equation}
B_{0} = \frac{\alpha^{3}}{2\pi} \frac{\Phi_{P}(R_{\star})R_{\star}}{\left[ \sin(\alpha R_{\star}) - \alpha R_{\star} \cos(\alpha R_{\star})\right]} \; ,
\label{eq:spheromakaxialfield}
\end{equation}

where $R_{\star}$ is the distance from the center of the spheromak where the magnetic field becomes completely axial.

\section{Comparison of the GCS and FRi3D White-Light Fitting Tools} \label{app:fri3d}

Both the GCS and FRi3D models can be used to visually fit a parameterized CME morphology to white-light coronagraph images from one or more viewpoints. The difference between the two methods resides in the number of free parameters that describe the CME shell---six for GCS and nine for FRi3D. The six GCS properties are: latitude, longitude, axial tilt, nose height, half-width, and aspect ratio (see also Table~\ref{tab:gcs}). The nine FRi3D properties, on the other hand, are: latitude, longitude, axial tilt, nose height, half-width, half-height, flattening, pancaking, and skew. It follows that the FRi3D model allows for much greater control and fine-tuning of the overall flux rope geometry, which is particularly useful in the case of strongly distorted and/or asymmetric CMEs.

\begin{figure*}[!t]
\center
	\includegraphics[width=0.99\linewidth]{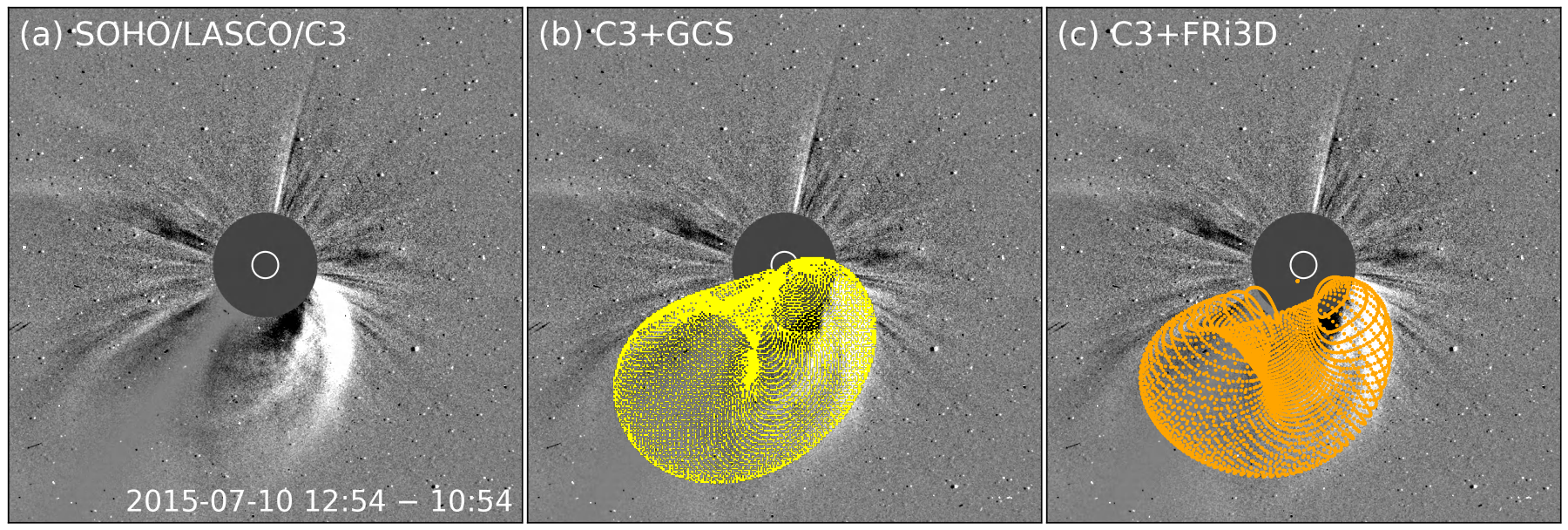}
	\caption{Comparison of a (b) GCS versus a (c) FRi3D fitting of the 2017 July 9 CME applied to (a) SOHO/LASCO/C3 imagery from July 10 at 12:54~UT.}
	\label{fig:gcs+fri3d}
\end{figure*}

Figure~\ref{fig:gcs+fri3d} shows a comparison of the GCS and FRi3D fits to the central portion of the extended 2015 July 9 CME---i.e., that displayed in Figure~\ref{fig:gcs_fits}(b) and corresponding to Part~2 in Table~\ref{tab:gcs}. It is evident that there are stark differences between the two models: The FRi3D reconstruction allows for a lower latitude of the flux rope axis, a tilt that is closer to the equatorial plane, and a clearer separation of the CME legs, while maintaining the overall appearance of east--west asymmetry (cf.\ also the EUHFORIA input parameters in Table~\ref{tab:euhforia} resulting from the two sets of fitting procedures).

%%%%% FIGURE A3 %%%%%
%%%%%%%%%%%%%%%%%%%

% From \citet{maharana2022}:

% \begin{equation}
% h_{P} = h_{T} \tan R_\mathrm{min} 
% \label{eq:rp}
% \end{equation}

% \begin{equation}
% h_{T} + h_{P} = 21.5\, R_{\odot}
% \label{eq:leadingedge}
% \end{equation}

%\begin{thebibliography}{}
\bibliographystyle{aasjournal}
\bibliography{bibliography} 
%\end{thebibliography}

\end{document}